\documentclass[11pt]{article}
\usepackage{amsmath,amssymb,color}

\usepackage{amsfonts}

\newcommand{\hoch}[1]{$\, ^{#1}$}

\newcommand{\be}{\begin{equation}}
\newcommand{\ee}{\end{equation}}
\newcommand{\bea}{\setlength\arraycolsep{2pt} \begin{eqnarray}}
\newcommand{\eea}{\end{eqnarray}}
\newcommand{\nn}{\nonumber}

\def\ft#1#2{{\textstyle{\frac{\scriptstyle #1}{\scriptstyle #2} } }}
\def\fft#1#2{{\frac{#1}{#2}}}

\def\0{{\sst{(0)}}}
\def\1{{\sst{(1)}}}
\def\2{{\sst{(2)}}}
\def\3{{\sst{(3)}}}
\def\4{{\sst{(4)}}}
\def\5{{\sst{(5)}}}
\def\6{{\sst{(6)}}}
\def\7{{\sst{(7)}}}
\def\8{{\sst{(8)}}}
\def\sst#1{{\scriptscriptstyle #1}}

\def\ep{{\epsilon}}
\def\del{{\partial}}
\def\im{{{\rm i}}}

\begin{document}

\begin{flushright}
\hfill{MIFPA 12-24}
\end{flushright}

\vspace{25pt}
\begin{center}
{\Large {\bf Gyrating Schr\"odinger Geometries and Non-Relativistic Field Theories}}

\vspace{10pt}

H. L\"u\hoch{1} and C.N. Pope\hoch{2,3}

\vspace{10pt}

\hoch{1}{\it Department of Physics, Beijing Normal University,
Beijing 100875, China}

\vspace{10pt}

\hoch{2} {\it George P. \& Cynthia Woods Mitchell  Institute
for Fundamental Physics and Astronomy,\\
Texas A\&M University, College Station, TX 77843, USA}

\vspace{10pt}

\hoch{3}{\it DAMTP, Centre for Mathematical Sciences,
 Cambridge University,\\  Wilberforce Road, Cambridge CB3 OWA, UK}

\vspace{40pt}

\underline{ABSTRACT}
\end{center}

We propose homogeneous metrics of Petrov type III that describe
gyrating Schr\"odinger geometries as duals to some non-relativistic
field theories, in which the Schr\"odinger symmetry is broken
further so that the phase space has a linear dependence of the
momentum in a selected direction. We show that such solutions can
arise in four-dimensional Einstein-Weyl supergravity as well as
higher-dimensional extended gravities with quadratic curvature terms
coupled to a massive vector. In Einstein-Weyl supergravity, the
gyrating Schr\"odinger solutions can be supersymmetric, preserving
$\ft14$ of the supersymmetry.  We obtain the exact Green function in
the phase space associated with a bulk free massive scalar.

\thispagestyle{empty}





\newpage

\section{Introduction}

  The AdS/CFT correspondence provides a way of relating
$d$-dimensional relativistic conformal field theories in the
strongly-coupled regime to bulk theories including gravity in $d+1$
dimensions.  It is also of considerable interest, in view of the
possible applications in condensed matter and other branches of
physics, to study situations where the boundary of the gravitational
theory does not possess the full relativistic symmetries of a
$d$-dimensional CFT.  For this reason, $(d+1)$-dimensional
gravitational backgrounds that are asymptotic not to anti-de Sitter
spacetime, but rather to so-called or Schr\"odinger
\cite{schr1,schr2} or Lifshitz spacetimes \cite{lifs1}, have been
considered.  In the Lifshitz case the boundary spacetime exhibits a
scaling behaviour in the time direction, characterised by the
Lifshitz exponent $z$, that differs from the $z=1$ scaling in the
spatial directions.  Thus when $z\ne 1$ there is an anisotropy
between the time direction and the spatial directions. In the
Schr\"odinger case the coordinate identified as time in the boundary
theory enters via only a first derivative when one considers wave
equations in the bulk background, thus leading to the expectation
that boundary theory will represent a quantum-mechanical system
described by the non-relativistic Schr\"odinger equation.

  A key ingredient in interpreting the nature of the boundary theory
is to study the symmetry group of the bulk system.  In the case of a
$(d+1)$-dimensional AdS bulk solution, this symmetry will be the
full conformal group $SO(d,2)$.  The asymptotic symmetry of a
Lifshitz or Schr\"odinger type spacetime will be reduced to some
subgroup of $SO(d,2)$.

   In this paper, we shall consider situations where the symmetry group
exhibits not only an anisotropy between time and the spatial
directions, but in addition an anisotropy within the spatial
directions themselves.  Such an anisotropy could arise, for example,
if there were a uniform electric or magnetic field along some
particular direction in the dual boundary system.  Our motivation
for investigating systems exhibiting a spatial anisotropy arose from
our finding that the associated bulk gravitational backgrounds turn
out to arise naturally in theories of gravity where the usual
Einstein-Hilbert action is augmented by higher-order curvature
contributions. The metrics we consider are of dimension $D=4+n$, and
take the form
\begin{equation}
ds^2 = \ell^2\left[\fft{dr^2 - 2 du dv + dx^2 + dy^i dy^i}{r^2} +
\fft{2 c_1\, du dx}{r^{z+1}} - \fft{c_2\,
du^2}{r^{2z}}\right]\,,\label{metricans}
\end{equation}
where $1\le i\le n$ and $c_i$ are constants. They have 1) shift, 2)
rotation, 3) boost and 4) dilatation symmetries given by
\begin{eqnarray}
1):&& \delta u=a^+\,,\qquad \delta v= a^-\,,\qquad \delta x=
a_x\,,\qquad \delta y_i= a_i\,;\cr 
2):&& \delta y_i=\Lambda_{ij}\, y_j\,,\qquad
\Lambda_{ij}=\Lambda_{[ij]}\,;\qquad 3):\qquad \delta y_i= b_i\,
u\,,\qquad \delta v = b_i\, y_i\,;\cr 
4):&& \delta x = w x\,,\qquad \delta y_i= w y_i\,,\qquad \delta r= w
r\,,\qquad \delta u = z w u\,,\qquad \delta v= (2-z) w
v\,.\label{symm}
\end{eqnarray}
The metrics are homogeneous, and all curvature invariants are
unchanged from their values in the AdS metrics that have
$c_1=c_2=0$.  If $c_1=0$, the metrics reduce to standard
Schr\"odinger metrics.  In general $c_i$ can be functions of all
coordinates except the null coordinate $v$.  The corresponding
metrics describe gyrating spacetimes. AdS gyratons were studied in
\cite{Frolov:2005ww}.

  We have investigated in some detail the case of four-dimensional
gravity with a quadratic curvature addition proportional to the
square of the Weyl tensor.  This theory, known as Einstein-Weyl
gravity, admits a supersymmetric extension to an off-shell theory of
supergravity. We find that for suitable choices of parameters our
solutions are supersymmetric. The four-dimensional metrics are  in
general of Petrov type III, degenerating to Petrov type N if $c_1=0$
or $z=1$ or $0$.

We also find solutions with spatial anisotropy in arbitrary higher
dimensions.  We have studied two different higher-derivative
theories in higher dimensions, namely Einstein gravity augmented by
the four-dimensional Gauss-Bonnet invariant, and Einstein gravity
instead augmented by the addition of the Weyl-squared invariant.

  Having obtained the gravitational backgrounds, we then consider a
minimally-coupled probe scalar field and compute the associated 2-point
functions in the boundary field theories.

\section{Einstein-Weyl Supergravity and BPS solutions}

The field content of the off-shell ${\cal N}=1$, $D=4$ supergravity
consists of the vielbein $e_{\mu}^a$, a vector $A_\mu$ and a complex
scalar $S + {\rm i} P$, totalling 12 off-shell degrees of freedom,
matching with that of the off-shell gravitino $\psi_\mu$. If one
just considers the supersymmetric extension of ordinary Einstein
gravity, then the fields $A_\mu$, $S$ and $P$ are auxiliary with
purely algebraic equations of motion \cite{sw,fv}. In the
supersymmetric extension of Einstein-Weyl gravity
\cite{LeDu:1997us,lpsw}, the field $A_\mu$ becomes a dynamical
massive vector, while $S$ and $P$ are still auxiliary. Adopting the
notation of \cite{lpsw}, the bosonic Lagrangian is given by
\begin{equation}
e^{-1}{\cal L} = R + \ft23 (A_\1^2 - S^2 - P^2) + 4S\,
\sqrt{-\Lambda/3} + \ft12\alpha C^{\mu\nu\rho\sigma}
C_{\mu\nu\rho\sigma} - \ft13\alpha F^{\mu\nu} F_{\mu\nu}\,,\label{genlag}
\end{equation}
where $C_{\mu\nu\rho\sigma}$ is the Weyl tensor and $F=dA$.
The supersymmetry transformation of the gravitino is
given by
\begin{equation}
\delta \psi_\mu = -D_{\mu} \epsilon - \ft{\rm i}6 (2A_\mu -
\Gamma_{\mu\nu} A^\nu)\Gamma_5 \epsilon - \ft16\Gamma_\mu (S + {\rm
i}\Gamma_5 P)\epsilon\,.\label{susytrans}
\end{equation}

The equations of motion for the scalar fields $S$ and $P$ imply that
$ S=3\sqrt{-\Lambda/3}$ and $P=0$. The vector equation of motion describes
a massive Proca field: $\alpha\,\nabla^{\mu} F_{\mu\nu} + A_\nu =0$.
The Einstein equation is
\begin{equation}
R_{\mu\nu}- \ft12 R g_{\mu\nu} + \Lambda\, g_{\mu\nu} -2 \alpha
E_{\mu\nu} = -\ft23 (A_\mu A_\nu - \ft12 A^2 g_{\mu\nu})
+\ft23\alpha (F_{\mu\nu}^2 - \ft14 F^2 g_{\mu\nu})\,,\nn
\end{equation}
where $E_{\mu\nu}= (\nabla^\rho\nabla^\sigma + \ft12
          R^{\rho\sigma})C_{\mu\rho\sigma\nu}$ is the Bach tensor.
It was shown in \cite{lpsw} that the theory admits a supersymmetric
AdS vacuum with cosmological constant $\Lambda$, and the linearised
spectrum of fluctuations around the AdS background was analyzed.
The gravity modes are
identical to those in Einstein-Weyl gravity, which was studied in
\cite{lpcritical}. There is a ghostlike massive
spin-2 mode in additional to the massless graviton.  The mass is
determined by the product $\alpha \Lambda$.
A special case, known as {\it Critical Gravity}, in which the massive
mode is replaced by a mode with logarithmic fall-off,
arises if $\alpha \Lambda = \ft32$.

   We consider a metric of the form (\ref{metricans}) in $d=4$,
with the massive vector $A_\mu$ given by $ A = q r^{-z} du$, where
$q$ is a constant.  (The form of the $r$-dependence here is dictated
by requiring invariance under the dilatation symmetry in
(\ref{symm}).)  We obtain solutions provided that
\begin{equation}
\alpha = -\fft{\ell^2}{z(z+1)}\,,\qquad
\Lambda=-\fft{3}{\ell^2}\,,\qquad q=\pm \ft32 (z-1) \sqrt{c_1^2 + 2c_2}\,.
\label{solution}
\end{equation}
The solutions with $c_1=0$ were obtained in \cite{ll}.

\medskip
\noindent
{\bf Supersymmetry analysis}
\medskip

Setting the AdS scale parameter $\ell=1$ for convenience, we
choose the vielbein basis
\be
e^+ = du\,,\qquad e^-=\fft{dv}{r^2} -\fft{c_1\, dx}{r^{z+1}} +
  \fft{c_2\, du}{2 r^{2z}}\,,\qquad e^2 = \fft{dx}{r}\,,\qquad
e^3= \fft{dr}{r}\,,
\ee
such that $ds^2 = \eta_{ab}\, e^a\otimes e^b$ with $\eta_{+-}=-1$,
$\eta_{22}=\eta_{33}=1$.  The inverse vielbein is given by
\be
E_+=\fft{\del}{\del u} - \ft12 c_2\, r^{2-2z}\, \fft{\del}{\del v}\,,\quad
E_-= r^2\, \fft{\del}{\del v}\,,\quad E_2 = r\fft{\del}{\del x} +
c_1\, r^{2-z}\, \fft{\del}{\del v}\,,\quad E_3 = r\fft{\del}{\del r}\,.
\ee
The Lorentz-covariant exterior derivative
$\nabla=d + \ft14 \omega^{ab}\, \Gamma_{ab}$ acting on spinors is then
given, in terms of its vielbein components, by
\bea
\nabla_+&=& E_+ -\ft12\, \Gamma_{+3} -\ft14 c_1 (z-1) r^{-z}\,\Gamma_{23}
 - \ft12 c_2 (z-1) r^{-2z} \,\Gamma_{-3}\,,\qquad
\nabla_- = E_- -\ft12 \,\Gamma_{-3}\,,\cr 
\nabla_2&=& E_2 -\ft12\,\Gamma_{23}+
  \ft14 c_1 (z-1) r^{-z} \, \Gamma_{-3}\,,\qquad
\nabla_3= E_3 -\ft14 c_1 (z-1) r^{-z} \,\Gamma_{-2}\,. \eea

  Requiring that the supersymmetry variation of the gravitino, given
by (\ref{susytrans}), vanish, we find that the solutions
(\ref{solution}) are supersymmetric if any of the following holds:
\bea
&&1)\quad c_1=0\,,\quad q=0:\quad
\ep=r^{-1/2}\, \ep_-\,,\quad \hbox{where}\quad
\Gamma_3\,\ep_-=\ep_-\,,\quad \Gamma_-\,\ep_-=0\,,\\
&&2)\quad c_2=0\,,\quad q=\ft32 c_1(z-1):\quad
  \ep=r^{-1/2}\, \ep_+\,,\quad \hbox{where} \quad
\Gamma_3\,\ep_+=\ep_+\,,\quad \Gamma_+\,\ep_+=0\,,\nn\\
&&3)\quad c_2=-\ft49 c_1^2\,,\quad q=-\ft12 c_1 (z-1):\quad
\ep=r^{-1/2}\, \ep_-\,,\quad \hbox{where}\quad
\Gamma_3\, \ep_-=\ep_-\,,\quad \Gamma_-\,\ep_-=0\,.\nn
\eea
In each case the projection conditions imply that the spinor
$\ep_\pm$, which is constant, is unique up to scaling.

More general supersymmetric gyraton solutions can also arise.
As we have mentioned earlier, in general the constants $c_i$ can be replaced
by functions of all coordinates except $v$ in gyrating geometries, and
so one may consider metrics of the form
\be
ds^2 = \fft{dr^2 -2 du dv + dx^2}{r^2} + f\, du dx + h du^2\,,
\qquad A=\phi\, du
\ee
where $f$, $h$ and $\phi$ are functions of $u$, $x$ and $r$.  Here, we
shall restrict our attention to the case where these functions depend only
on $r$.  The general such bosonic solution can easily be obtained explicitly.
Since it is a little complicated to present, we shall not give it here.
If in addition we require supersymmetry, then we find that the Killing
spinor must satisfy the projections $\Gamma_3\ep=\ep$ together with
either $\Gamma_+\ep=0$ or $\Gamma_-\ep=0$.  In these two cases, the
$r$-dependent functions must satisfy
\be \Gamma_+\ep=0:~ \phi= -\ft32 r\, (f + \ft12 r\, f')\,,\quad h +
\ft12 r\, h'=0\,;\qquad \Gamma_-\ep=0:~ \phi= \ft12 r\, (f + \ft12
r\, f')\,. \ee
For generic values of $z$, related to $\alpha$ by $\alpha=-1/(z(z+1))$,
imposing these supersymmetry restrictions on the general bosonic solution,
and discarding trivial terms that can be immediately removed by coordinate
transformations, we find:
\bea
\Gamma_+\ep=0:&& f= \fft{a_1}{r^{1+z}} + a_2\, r^z\,,\qquad
 \phi = \fft{3(z-1)\, a_1}{4 r^z} -\fft{3 (z+2) a_2\, r^{z+1}}{4}\,,\qquad
h=0\,,\\
\Gamma_-\ep=0:&& f= \fft{a_1}{r^{1+z}} + a_2\, r^z\,,\qquad
  \phi = -\fft{(z-1)\, a_1}{4 r^z} +\fft{(z+2) a_2\, r^{z+1}}{4}\,,\nn\\
&& h= \fft{a_1^2}{9 r^{2z}} + \ft19 a_2^2\, r^{2z} +
\fft{b_1}{r^{z+1}} + b_2\, r^z + b_3\, r\,. \eea
A special case arises in the case of critical gravity, where
$\alpha=-\ft12$, i.e.~where $z=1$ or $z=-2$.  We find the
supersymmetric solutions with logarithmic fall-off
\bea
\Gamma_+\ep=0:&& f= \fft{a_1\, \log r}{r^2} + a_2\, r\,,\qquad
\phi= -\fft{3 a_1}{4 r} -\ft94 a_2\, r^2\,,\qquad h=0\,,\\
\Gamma_-\ep=0:&& f= \fft{a_1\, \log r}{r^2} + a_2\, r\,,\qquad
  \phi= \fft{a_1}{4r} + \ft34 a_2\, r^2\,,\nn\\
&&h=\fft{a_1^2}{27 r^2}(2+4\log r +3 \log^2 r) + \ft19 a_2^2\, r^4 +
\fft{b_1\, \log r}{r^2} + b_2\, r + b_3\, r\log r\,.
\eea

\section{Generalization to Higher Dimensions}

Solutions in higher dimensions $d\ge5$ can also arise. We find that
Einstein gravity with additional quadratic curvature terms, coupled
to a massive vector field, can support solutions with non-vanishing
$c_1$ for general $z$.  We consider the Lagrangian
\begin{equation}
e^{-1}{\cal L}_d = R - 2\Lambda + \alpha R^2 + \beta
R^{\mu\nu} R_{\mu\nu} + \gamma E_{GB}^2 - \ft14 F^2 - \ft12 \sigma^2 A^2\,.
\end{equation}
The equations of motion for the vector $A_\mu$ imply that $\sigma =
z(z+d-3)/\ell^2$. There are then three further equations from the
variation of $g_{\mu\nu}$, and so with as-yet unspecified
coefficients $(\alpha, \beta \,\gamma,\sigma,\Lambda)$ in the
theory, solutions must exist.

As an example, consider the case when $\alpha=\beta=0$, corresponding
to Gauss-Bonnet, or Lovelock, gravity.  We then find
\begin{equation}
\Lambda = - \fft{(d-1)(d-2)}{4\ell^2}\,,\qquad \gamma =
\fft{\ell^2}{2(d-3)(d-4)}\,,\qquad q^2 = -\fft{(z-1)^2 c_1^2
\ell^2}{4\ell^2}\,.
\end{equation}
Note that $c_2$ is arbitrary in this case.  The fact that $q^2$ is negative
implies that we should really send $A_\mu\rightarrow \im\, A_\mu$,
implying that the massive vector is ghostlike.

As another example, we may consider Einstein-Weyl gravity in $d$
dimensions, corresponding to taking
\begin{equation}
\alpha = \fft{4(d-3)}{d-2}\gamma\,,\qquad \beta = - \fft{d(d-3)}{(d-1)(d-2)}\,.
\end{equation}
We then find that
\begin{eqnarray}
&&\Lambda = -\fft{(d-1)(d-2)}{2\ell^2}\,,\qquad \gamma = -
\fft{(d-2)\ell^2}{4(d-3)z(z+d-3)}\,,\cr && q^2 =
-\fft{\ell^2(z-1)^2}{z^2(z+d-3)}\Big( 2 (d-4 + 3z) c_2 + \Big(\ft34
z + \fft{(d-4)^2}{4(d-3)}\Big) c_1^2\Big)\,.
\end{eqnarray}
In the next section we shall require that $c_1^2 + c_2>0$.  For suitable
allowed choices of $c_1$ and $c_2$, $q^2$ can be real in this case.

\section{Boundary Field Theory}

   We consider a scalar field $\Phi$ with mass $m_0$
that is minimally coupled to the
background metric (\ref{metricans}).  With
\be
\Phi= f(r)\, e^{-\im\omega t-\im p \, v}\,
e^{\im\vec k\cdot\vec y + \im k_x x}\,,
\ee
where $u=t$ is taken to be the time coordinate,
the bulk wave equation $(\square -m_0^2)\Phi=0$
gives
\bea
f'' -\fft{(n+2)}{r}\, f' +\Big[ -\fft{m_0^2}{r^2} -
  \fft{(c_1^2+c_2)p^2}{r^{2z-2}} +\fft{2 c_1 p\, k_x}{r^{z-1}} -
\kappa^2\Big]f=0\,,\label{wavez}
\eea
where
\be
\kappa^2 = \vec k^2 + k_x^2 + 2 \omega\,p\,.
\ee
If one assumes that the coordinate $v$ is compactified on a circle,
then $p$ will be quantised. Solutions of the form $f\sim r^\Delta$
near the boundary at $r=0$ will exist if $z\le 2$, and at the
limiting value $z=2$ equation (\ref{wavez}) becomes
\bea
f'' -\fft{(n+2)}{r}\, f' +\Big[ -\fft{m^2}{r^2}
+\fft{2 c_1 p k_x}{r} -
\kappa^2\Big]f=0\,,\label{wavecrit}
\eea
where
\be
m^2= m_0^2 + (c_1^2 + c_2)\, p^2\,.
\ee
It is necessary that $c_1^2 + c_2$ be non-negative in order that
$m^2$ be positive for all values of the (quantised) $v$-momentum
$p$. The asymptotic forms of the solutions at small $r$ are $f\sim
r^{\Delta_\pm}$ with
\be
\Delta_\pm= \ft12(n+3) \pm \nu\,,\qquad
\nu=\ft12 \sqrt{(n+3)^2 + 4 m^2}\,.
\ee
The exact form of the general solution is
\be
f= \alpha_1\, r^{\Delta_+}\, M(a,b;2\kappa r)\, e^{-\kappa r} +
   \alpha_2\, r^{\Delta_+}\, U(a,b;2\kappa r)\, e^{-\kappa r}\,,
\ee
where $M(a,b;2\kappa r)$ and $U(a,b;2\kappa r)$ are the confluent
hypergeometric functions of the first and second kinds, and we have
defined
\be
a=\nu+\ft12 -\fft{c_1\,p\, k_x}{\kappa}\,,\qquad b=2\nu+1\,.
\ee

  Demanding normalisability at large $r$ requires that we reject the
exponentially diverging solution involving $M(a,b;2\kappa r)$, giving
\be
f= r^{\Delta_+ r}\, U(a,b;2\kappa r)\, e^{-\kappa r}\,,
\ee
which decays exponentially at infinity. At small $r$ we use the relation
\be
U(a,b;2\kappa r) = \fft{\Gamma(1-b)}{\Gamma(a-b+1)}\,
 M(a,b;2\kappa r) + \fft{\Gamma(b-1)}{\Gamma(a)}\, (2\kappa r)^{1-b}\,
 M(a-b+1,2-b;2\kappa r)\,,
\ee
which shows, since $M(a,b;2\kappa r)$ is analytic at small $r$, that aside
from contact terms the leading-order form for $f$ is
\be
f\sim (\kappa r)^{\Delta_-}\, \Big[ 1 +
  \fft{\Gamma(a)\Gamma(1-b)}{\Gamma(a-b+1)\Gamma(b-1)}\, r^{2\nu}\Big]\,.
\ee
Thus, using the prescription \cite{sonsta}
\be
G_R(k) = 2{\cal F}(k,r)|_{r=r_B} \sim \sqrt{-g} g^{rr}\, \del_r\log f
\ee
for the retarded Green function, we find
\be
{\cal F}(k,r)|_{r=\ep} \sim -\fft{(\ep\kappa)^{2\nu}}{\ep^{n+3}}\,
\fft{\Gamma(1-2\nu)}{\Gamma(2\nu)} \,
\fft{\Gamma(\nu+\ft12 -c_1\, p \, k_x/\kappa)}{
           \Gamma(-\nu- \ft12 +c_1\, p \, k_x/\kappa)}\,.
\ee
Unlike the situation in a pure Schr\"odinger spacetime background,
where momentum dependence of the Green function enters only via the
$(\ep\kappa)^{2\nu}$ factor, here when $c_1\ne0$ there is momentum
dependence in the Gamma functions also.

\section{Conclusions}

In this paper, we have investigated certain four-dimensional
generalisations of Schr\"odinger metrics that arise naturally as
solutions of Einstein-Weyl gravity.  Their symmetry group is smaller
than that of the Schr\"odinger metrics, correspoding to an
inisotropy not only between space and time, but also among the
spatial coordinates themselves.  The metrics are of Petrov type III,
reducing to type N in the Schr\"odinger limit.  Einstein-Weyl
gravity can be viewed as the bosonic sector of an off-shell ${\cal
N}=1$ supergravity theory, and we find that included amongst the
solutions we have obtained are some that are supersymmetric. We also
considered higher-dimensional analogues of the four-dimensional
solutions, and showed that they can also arise in Einstein-Weyl and
in Einstein-Lovelock gravity in arbitrary dimensions $d>4$.

The solutions we have obtained provide natural backgrounds for
studying the dual strongly-coupled non-relativistic boundary
theories.  The spatial anisotropy would correspond to some breaking
of rotational symmetry in the boundary theory, such as might arise
from a uniform electric or magnetic field.  We calculated the
two-point correlation function for boundary operators dual to a
minimally-coupled massive scalar probe field in the bulk theory.

Higher-order off-shell supergravities have rich structures for
constructing geometries not only for relativistic but also for
non-relativistic field theories, in the context of supersymmetry.
It would be also of great interest to investigate whether there exist
two-derivative gravity theories that also gives rise to gyrating
Schr\"odinger geometries.

\section*{Acknowledgements}

We are grateful to Sera Cremonini for useful discussions, and 
to the KITPC, Beijing, for hospitality during the
course of this work. The research of H.L.~is supported in part by
NSFC grant 11175269.  C.N.P. is supported in part by DOE grant
DE-FG03-95ER40917.

\end{document}